# City-level energy and emission assessment based on 20+ million electric vehicle registrations in China

Yanqiao Deng [1, 2], Minda Ma [3] *, Nan Zhou [4] *, Hong Yuan [1], Zhili Ma [1], Xin Ma [5]

1. School of Future Technology, South China University of Technology, Panyu District, Guangzhou 511442, China
2. School of Management Science and Real Estate, Chongqing University, Chongqing 400045, PR China
3. School of Architecture and Urban Planning, Chongqing University, Chongqing, 400045, PR China
4. Energy and Resources Group, University of California, Berkeley, CA 94720, United States
5. School of Mathematics and Physics, Southwest University of Science and Technology, Mianyang, 621010, PR China

- Corresponding author: Dr. Minda Ma, Email: minda.ma@cqu.edu.cn
  Homepage: https://globe2060.org/MindaMa/
  Corresponding author: Dr. Nan Zhou, Email: nzhou@berkeley.edu
  Homepage: https://buildings.lbl.gov/people/nan-zhou

- Lead contact:
  Dr. Minda Ma, Email: minda.ma@cqu.edu.cn

## GRAPHICAL ABSTRACT

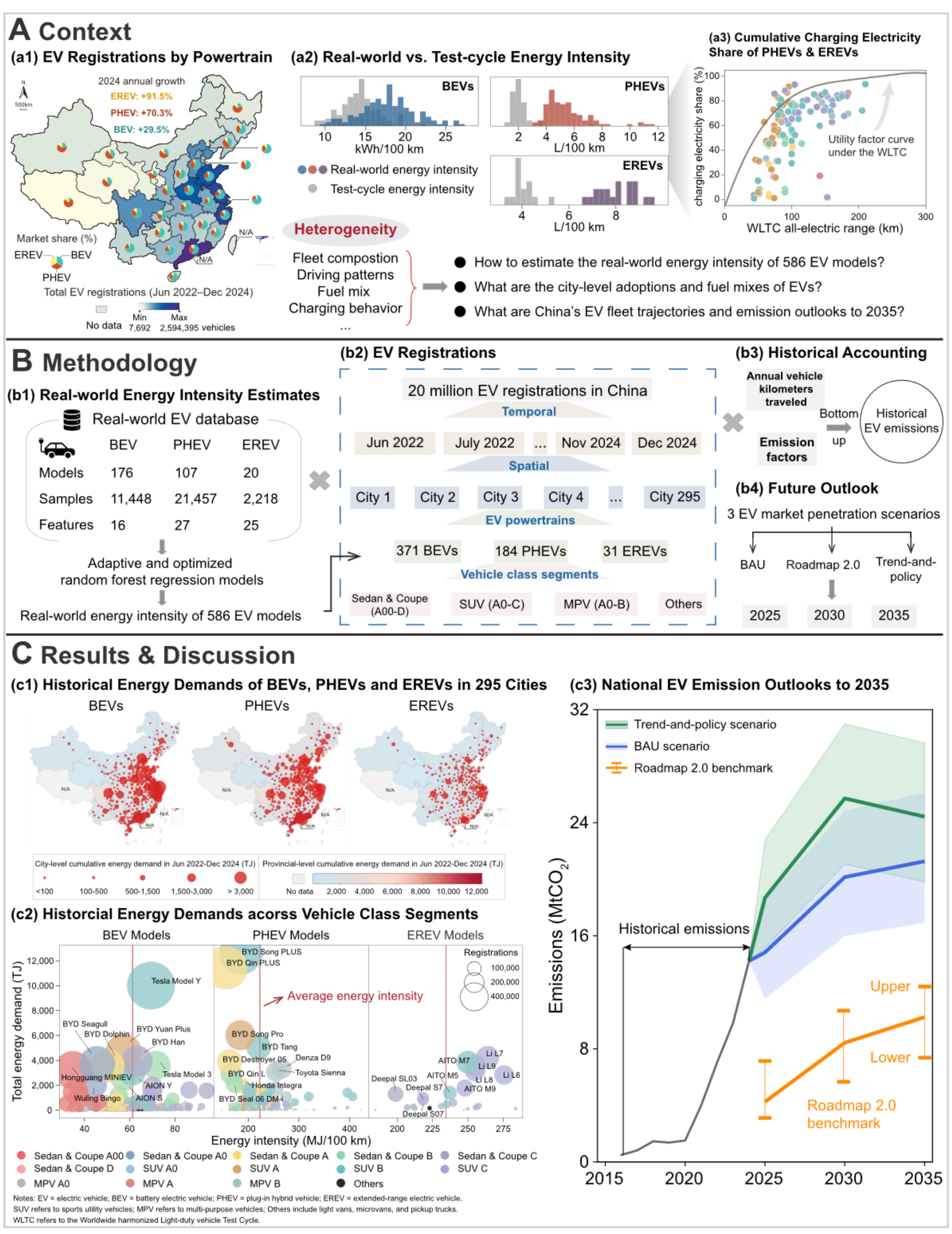

**Graphical abstract.** (A) Heterogeneity in (a1) EV registrations, total and percent of different powertrains, (a2) differences between real-world and test-cycle energy intensities, and (a3) differences in charging characteristics of PHEVs and EREVs. (B) Refined bottom-up framework integrating (b1) real-world energy intensity estimates of 586 EV models via data-driven random forest regression for (b2) more than 20 million EV registrations at a monthly resolution to derive (b3) historical well-to-wheel energy use and $CO_2$ emissions and (b4) future scenario outlooks. (C) Historical energy demand of BEVs, PHEVs, and EREVs (c1) in 295 cities and (c2) within vehicle class segments during 2022–2024, and (c3) nationwide EV emission outlooks to 2035 under three scenarios.

## HIGHLIGHTS

- First city-level assessment of energy intensity of EVs in 295 cities of China in 2022–2024.
- Real-world intensities exceeded test-cycle values by 29% to 137% for 586 EV models.
- Gasoline supplied about 44% of EV energy demand despite rapid electrification.
- National emissions will peak in ~2030 and decline given adoption of solid-state batteries.

## eTOC

China's EV boom is furthering decarbonization of the world's transport sector, yet real-world efficiencies and emissions differ greatly among regions. Utilizing more than 20 million registrations of 586 EV models in 295 cities from 2022 to 2024, Deng et al. construct the first real-world, high-resolution database of China's EV market and project transition pathways to 2035. The results reveal real-world energy intensities that are 29% to 137% higher than test-cycle values and a 44% reliance on gasoline fuel despite rapid electrification. The study provides a robust data foundation and insights to evaluate and guide low-carbon EV transitions worldwide.

## ABSTRACT

China, the world's largest electric vehicle (EV) market, plays a pivotal role in global decarbonization of the transport sector. We present the first high-resolution assessment of EV adoption in 295 cities, utilizing more than 20 million registrations of 586 EV models tracked monthly from 2022 to 2024 and projecting transition pathways to 2035. Real-world data reveal that EVs are 30.9–212.8 megajoules per 100 km more energy efficient than internal combustion vehicles, yet their carbon intensities range from 18.2 to 270.4 g$CO_2$/km among provinces. The limited electrification of hybrids means that gasoline still accounts for 44% of EV energy use. Scenario projections suggest that emissions will peak about 2030 at 21.1–30.9 megatonnes of $CO_2$ and decline by 2035 under continued market transition. The findings establish an empirical foundation for accurate emissions accounting, emphasize the need to reduce regional disparities in adoptability, and offer globally relevant insights for road-transport decarbonization.

## BROADER CONTEXT

China is the world's largest EV market, where EV sales surpassed 11 million in 2024, making China integral to the global decarbonization of the transport sector. Yet real-world EV energy use and emissions often deviate from test standards, and the pace of low-carbon dispersion differs significantly among regions. Understanding the real-world dynamics of EV adoption and performance is crucial for designing effective policies to meet China's carbon neutrality goals and to inform transport transitions globally. In this study, we construct the first nationwide, city-level EV emission database at a monthly resolution, covering June 2022 to December 2024. Drawing on more than 20 million registrations of 586 EV models in 295 cities and a unified regression model trained on more than 34,000 samples, we capture real-world energy intensities and their regional disparities. Our results show that although EVs remain substantially more efficient than internal combustion engine vehicles, their energy intensities exceed test-cycle values for all powertrains. Additionally, carbon intensities differ widely among provinces and cities—shaped by fleet composition, grid carbon intensity, and charging behavior. Our findings provide an empirical foundation for robust accounting of EV emissions in China, highlight the urgency of addressing regional disparities, and deliver globally relevant insights for accelerating transitions to low-carbon transport sectors.

## INTRODUCTION

### Background

China already has emerged as the world's largest electric vehicle (EV) market,[1] where sales surpassed 11 million in 2024—a nearly 40% year-over-year (YoY) increase and more than triple the combined sales of EVs in Europe and the United States.[2] Although the United States recently has downplayed its carbon neutrality ambitions[3] and Europe is slowing its net-zero transition,[4] China continues to lead the global EV transition. Beyond sheer scale, China's market features exceptional diversity,[5] more than 580 distinct models having been introduced between mid-2022 and late 2024, accounting for more than 70% of the nearly 780 EV models available worldwide.[2] The combination of market scale, model diversity, and rapid urban adoption[6] makes China an ideal case for assessing real-world EV performance and decarbonization potential, providing useful insights for understanding global pathways to EV decarbonization.

A vital data gap remains: high-resolution, city-level data on the array of EVs, including battery electric vehicles (BEVs), plug-in hybrid electric vehicles (PHEVs), and extended-range electric vehicles (EREVs) for all vehicle classes under real-world operations that consider heterogeneity in fleet composition, grid carbon intensity, charging behavior, fuel mix, and driving patterns.[7] Conventional emissions accounting methods, based on standardized test cycles, generally fail to capture the complexity of real-world driving.[8] In particular, variable owner charging behaviors[9] cause the mixed-mode energy use of PHEVs and EREVs to deviate from theoretical assumptions, complicating robust estimation and making real-world simulations both challenging and costly.[10] A unified framework for quantifying energy demand and emissions based on real-world carbon intensity is essential to provide an empirically grounded basis for identifying pathways to passenger vehicle decarbonization in China and for informing market design, technical guidance, and policy formulation elsewhere.

### Literature review and research gap

When assessing well-to-wheel (WTW) emissions of EVs, conventional top-down approaches based on statistical yearbooks or simulations of a few EV models incorporate systematic errors and fail to capture the diversity of China's EV market and the heterogeneity among powertrains.[11,12] Recent efforts have turned to bottom-up frameworks[13] based on established benchmarks[14] and widely used transport-emission models[a].[18] The accuracy of those models as applied to China is limited, however, because they rely largely on standardized test-cycle data, static parameters, or inadequate scenario assumptions, overlooking dynamic factors such as driving patterns, charging behavior, and real-world energy efficiency.[19,20]

Although recent studies have incorporated environmental factors (e.g., temperature[21] and road conditions[22]) and the effects of charging infrastructure,[23] most analyses focus on the provincial scale,[24] concentrate on a single type of powertrain, and rely on outdated datasets, the most recent comprehensive data collection having stopped in 2023.[25] City-level assessments that reflect more recent real-world energy use by a range of BEV, PHEV, and EREV models are lacking. This gap limits our ability to track real-world energy trends and accurately assess emissions throughout a country. The information gaps must be addressed in order to establish real-world historical benchmarks and enable data-driven outlooks for future distribution of passenger EVs.

**Objectives and novelty**

To address key data gaps in current emissions accounting and to support comprehensive assessment of China's EV transition, we develop a monthly, city-level database of real-world energy demand and emissions that includes BEVs, PHEVs, and EREVs. Through a modified bottom-up WTW emissions accounting framework combined with real-world estimates of energy intensity, we quantify the energy demand and emissions of China's EV fleet from 2022 to 2024 and project transition pathways to 2035 under three market-penetration scenarios. Specifically, we raise three key questions.

- What are the real-world energy intensity distributions across vehicle powertrains and

---

[a] These models include the *Computer Programme to Calculate Emissions from Road Transport* (COPERT), developed by the European Environment Agency;[15] the *MOtor Vehicle Emission* Simulator (MOVES), developed by the U.S. Environmental Protection Agency;[16] and the *Greenhouse Gases, Regulated Emissions, and Energy Use in Transportation* (GREET) model, developed at Argonne National Laboratory.[17]

class segments?

- What is the city-level status of EV decarbonization shaped by charging behavior?
- What are the historical trajectories and emission outlooks for China's EV fleet to 2035?

**To address the above questions, we evaluate the city-level WTW energy demand and carbon emissions of passenger EVs in China from June 2022 to December 2024 and project the national outlook to 2035.** For the first question, we apply a unified data-driven regression model trained on more than 34,000 empirical samples to estimate real-world energy intensities among 586 BEV, PHEV, and EREV vehicle models, establishing a robust real-world baseline for assessing energy demand and emissions. For the second question, we modify the bottom-up framework by integrating monthly city-level EV registrations, provincial grid carbon intensities, real-world estimates of energy intensity, and annual vehicle kilometers traveled. Specifically, the framework separately quantifies electricity- and gasoline-based demand and emissions for PHEVs and EREVs, incorporating charging behavior via cumulative percentages of charging electricity instead of relying on a theoretical utility factor (UF). For the third question, we extend the analysis to 2035 by projecting transition pathways under three market-penetration scenarios, evaluating fleet composition, market evolution, and emission peaks under different policy and technical trajectories. Full methodological details are provided in the Experimental Procedures section of this paper and of the Supplemental Information.

**This study's most important innovation lies in being the first systematic evaluation of China's current EV market and real-world decarbonization trajectories for 295 cities, which is based on more than 20 million registrations of 586 EV models tracked monthly from 2022 to 2024. This high-resolution database provides the detailed historical data needed to project EV transition pathways to 2035 nationwide and supports future research on EV decarbonization pathways.** Overcoming the underestimation of test-cycle energy intensities, the regression model we developed enables the first comprehensive, real-world energy intensity database of China's EV market for all vehicle class segments. The model is applicable to multiple powertrains, captures multidimensional driving determinants through interpretability analysis, and demonstrates strong generalizability to unseen data, offering a scalable and reliable

foundation for large-scale EV emissions accounting. By integrating EV registration patterns with energy demand and emissions, our findings reveal pronounced regional differences in low-carbon development among provinces and cities, shaped by grid carbon intensity, EV adoption, and fleet composition. Notably, when charging behavior is incorporated, total gasoline consumption of PHEVs and EREVs approaches total electricity use, revealing that EVs have a higher-than-expected reliance on gasoline use. Building on historical datasets, our projections to 2035 underscore that targeted policy strategies at the city and provincial scales are important in the near term, and that full BEV electrification remains indispensable for thoroughgoing decarbonization. Beyond the aforementioned contributions, this study provides an important data foundation for future research and policy design for China's passenger transport decarbonization. It also offers globally relevant insights from the world's largest EV market to help guide transitions to low-carbon road transport worldwide.

## RESULTS

Our study identifies trends in EV registrations, estimates energy intensities of 586 EV models, estimates carbon intensities within provinces, and estimates the energy demands of 20 million EV registrations in 295 cities.

### Examination of EV registrations in China

We conducted an analysis of more than 20 million EV registrations in China from mid-2022 to the end of 2024, revealing a market characterized by pronounced heterogeneity within three key dimensions: geographical distribution, temporal growth patterns, and technological market segmentation (see Figure 1).

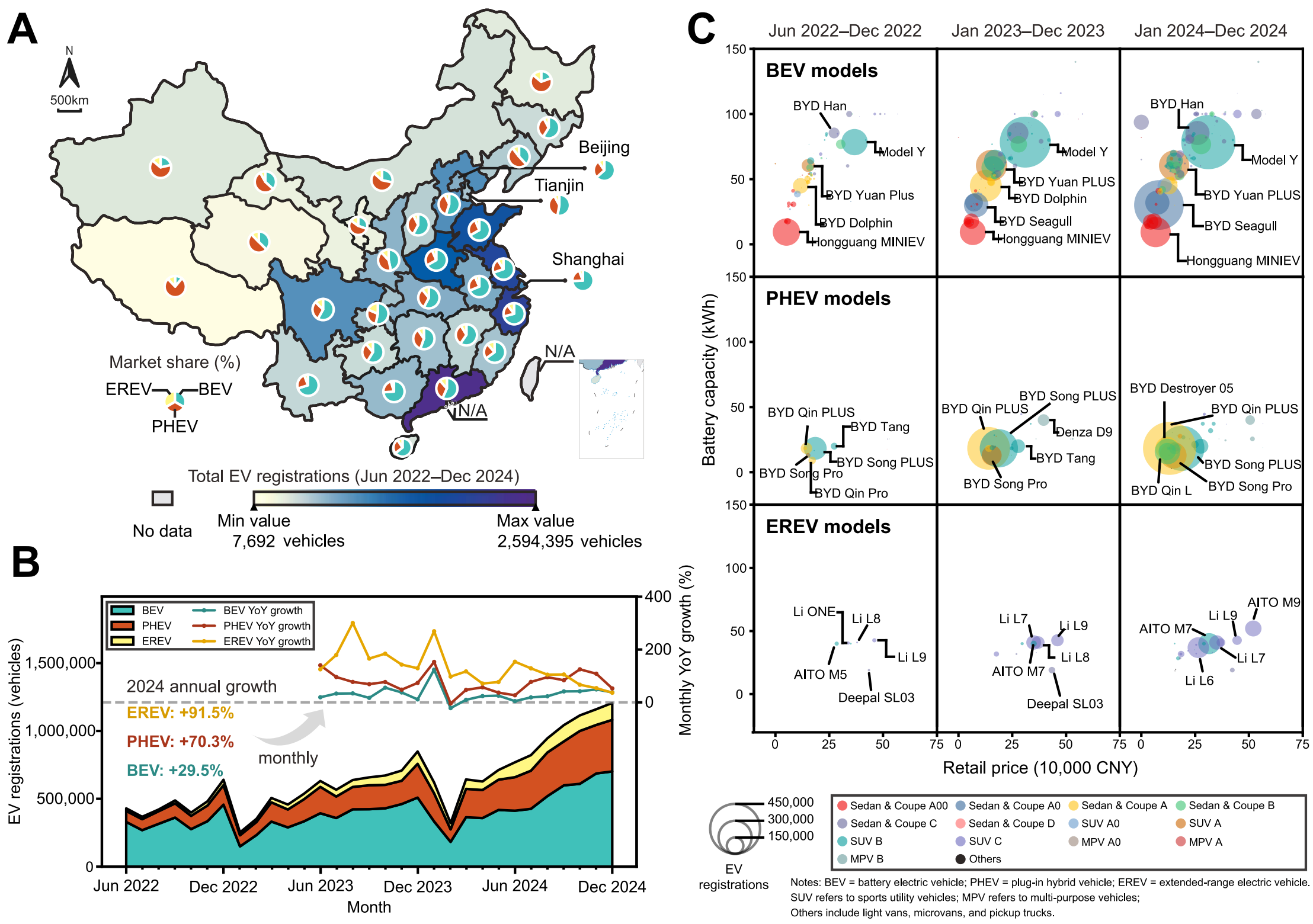

**Figure 1.** Distribution of EV registrations in China, Jun 2022–Dec 2024. (A) Provincial EV registrations and proportions of BEVs, PHEVs, and EREVs; (B) monthly registrations and YoY growth trends; and (C) registrations by vehicle class segments for various periods.

Regarding geographical distribution (see Figure 1A), EV registrations were concentrated in the eastern and southern coastal cities, which together accounted for approximately 68% of the national total (see Figure S1 of the Supplemental Information).

In contrast, the western and northeastern cities together represented less than 10% of the EV market. In the milder climates of the eastern and southern regions, BEVs constituted more than 60% of the local EV fleet. In northern provinces, however, where colder winters prevail, the BEV market fell below 40%, whereas PHEVs and EREVs gained market shares. This pattern is consistent with climate-related differences in consumer choices (see Figures S2 and S3), as hybrid powertrains mitigate the performance losses of batteries at low temperatures.

In terms of temporal growth (see Figure 1B), during the study period China's EV market continued to expand rapidly. Although BEVs consistently dominated total registrations, the percentages of PHEVs and EREVs rose rapidly. In 2024, the annual YoY growth in BEV registrations was 29.5%, compared with 70.3% for PHEVs and 91.5% for EREVs. Monthly YoY growth rates show similar differences, indicating that the market is electrifying rapidly but through two parallel pathways, battery electric and hybrid vehicles.

Regarding model diversity (see Figure 1C), by 2024 China's EV market encompassed more than 580 models, reflecting both rapid electrification and a flourishing industry. Of those models, 60% were fully electric, including 371 models led by BYD and Tesla. Consumer options included 184 PHEVs, largely from BYD, and 31 EREVs, primarily from Li Auto and AITO. Domestic manufacturers accounted for nearly 80% of all models, and registrations remained concentrated in a few "star" models.

**Real-world estimates of energy intensities**

Figure 2 illustrates the process we used to estimate real-world energy intensities using a data-driven regression framework to address missing data. Empirical samples from 176 BEV, 107 PHEV, and 20 EREV vehicle models (see Figure 2A) were used to train the random forest (RF) regression models. The trained models achieved high accuracy and stable performance (see Figure 2B). Predictions for vehicle models lacking empirical data were then incorporated into observed samples to generate the distribution for 586 vehicles total, 371 BEVs, 184 PHEVs, and 31 EREVs (see Figure 2C).

As shown in Figure 2A, the empirically derived real-world energy intensities revealed differences among the three EV powertrain types when standardized to megajoules (MJ)/100 kilometers (km). BEVs had the lowest median intensity at 61.0 MJ/100 km, whereas PHEVs and EREVs reached 215.6 and 242.9 MJ/100 km, respectively, approximately 3.5 to 4.0 times greater than that of BEVs. More detailed distributions by vehicle class segments are provided in Figure S4. As shown in the marginal histograms in Figure 2A, real-world energy intensity is systematically underestimated by official test-cycle values, most notably for EREVs (+137%), followed by PHEVs (+97.7%) and BEVs (+28.9%).

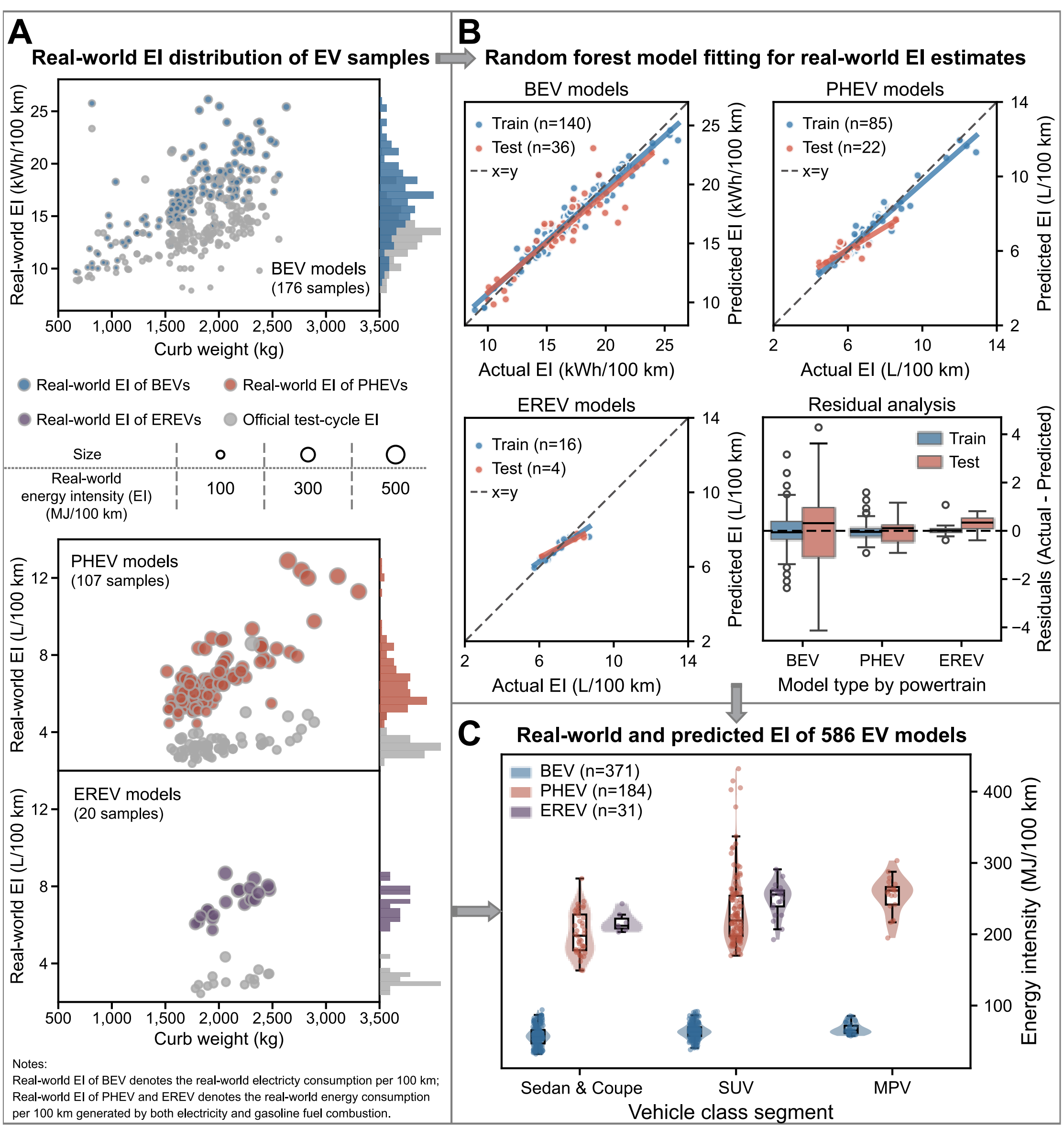

**Figure 2.** Real-world energy intensity (EI) distributions and estimates based on RF regression. (A) Empirically derived energy intensity distributions for samples of BEV, PHEV, and EREV models; (B) model fitting results via RF regression; and (C) overall energy intensity distributions

of all BEV, PHEV, and EREV class segments after combining empirical samples and RF model predictions.

The RF models were trained and validated separately for each powertrain. As shown in Table S1, the coefficient of determination ($R^2$) values reached 0.95 for BEVs and PHEVs and 0.83 for EREVs on the training samples. On the held-out test samples, the regression models retained strong generalizability, showing $R^2$ values of 0.78, 0.73, and 0.65 for the three powertrain types, respectively. As shown in Figure 2B, residuals were centered around zero with symmetric distributions, indicating no systematic bias. Segment-specific performance is further summarized in Table S2 and Figure S5. Model performance remained robust for BEVs and PHEVs across major vehicle class segments, with test-set $R^2$ values of 0.77 and 0.75 for BEV Sedan & Coupe and SUVs, respectively, and 0.73 and 0.55 for PHEV SUVs and Sedan & Coupe, respectively. In contrast, segment-specific inference for EREVs was more limited due to the small stratified sample size.

After integrating RF predictions with empirical data (see Figure 2C), the average real-world energy intensity was 61.2 MJ/100 km for BEVs, substantially lower than 227.9 MJ/100 km for PHEVs and 241.1 MJ/100 km for EREVs. Despite exhibiting higher real-world intensities than test-cycle values, all EV types outperformed internal combustion engine vehicles (ICEVs) relative to the International Energy Agency (IEA) benchmark[b] of 273.8 MJ/100 km. Specifically, BEVs consumed 212.6 MJ/100 km less, while PHEVs and EREVs consumed 32.7–45.9 MJ/100 km less.

### Estimated real-world carbon intensities

Based on the estimates of real-world energy intensities, we assessed carbon intensities per province and vehicle class segment (see Figure 3).

[b] IEA (2024). Global EV Outlook 2024. IEA, Paris. https://www.iea.org/reports/global-ev-outlook-2024.

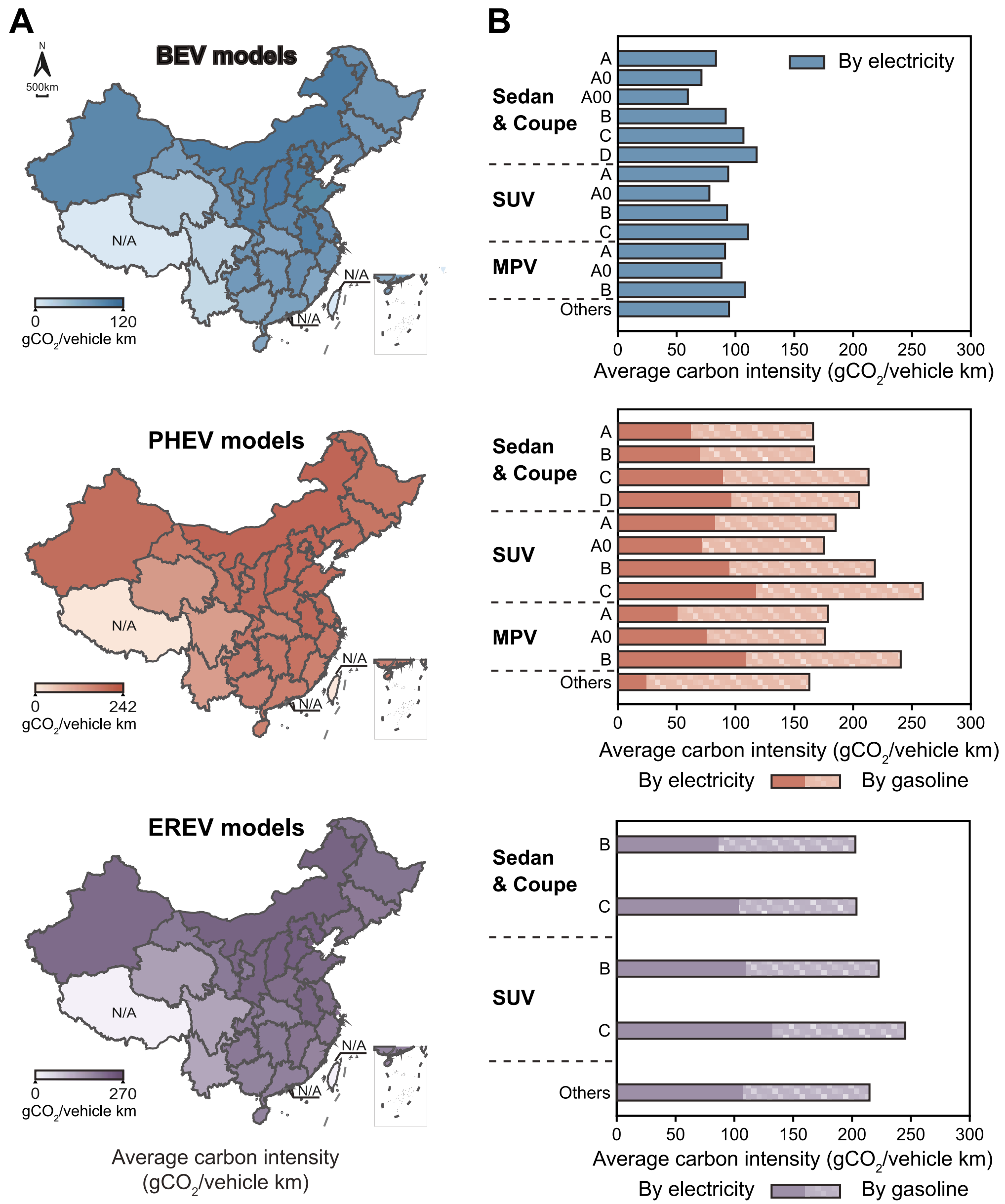

**Figure 3.** Real-world average carbon intensities of EVs in China (2022–2024). (A) Provincial heterogeneity of carbon intensities for BEVs, PHEVs, and EREVs; and (B) carbon intensities generated by electricity and gasoline among vehicle class segments. Note: “Others” shown in Panel B include light vans, microvans, and pickup trucks.

Figure 3A shows the average real-world carbon intensities across provinces from 2022 to 2024. Nationally, BEVs exhibited an average carbon intensity of 88.5 grams of carbon dioxide emitted per km traveled ($gCO_2$/km per vehicle), with values ranging from 18.2 to 119.7 $gCO_2$/km per vehicle—a sixfold variation. PHEVs averaged 207.6 $gCO_2$/km per vehicle, with a range of 136.6 to 242.4 $gCO_2$/km per vehicle, while EREVs averaged 226.0 $gCO_2$/km per vehicle. Combined with provincial EV registrations (see Figure S6 and Supplemental Note S1), these results highlight substantial regional disparities in EV decarbonization and underscore the need for province-specific strategies and cross-regional coordination.

Figure 3B presents a breakdown of real-world average carbon intensities by powertrain and vehicle class segments. Across all segments, BEVs exhibited the lowest intensities (59.5–117.9 $gCO_2$/km per vehicle), outperforming PHEVs by 67.6–147.5 $gCO_2$/km per vehicle and EREVs by 96.2–133.6 $gCO_2$/km per vehicle. Meanwhile, electricity accounted for only 30%–50% of the energy consumed by PHEVs and EREVs, indicating continued reliance on gasoline and substantial variation driven by driving conditions and charging behavior.[9]

**Energy demand of 20 million EV registrations in 295 cities**

City-level assessments reveal wide disparities in EV energy demand, underscoring the need for regionally aligned policies and stronger commitments from automakers to deliver genuine efficiency gains (see Figure 4).

Figure 4A shows that city-level EV energy demand during 2022–2024 was concentrated in economically dynamic regions. Guangdong led in consuming 28,187 terajoule (TJ), followed by Zhejiang and Jiangsu. In northern provinces such as Hebei and Liaoning, energy demand of PHEVs was 1.6 to 2.7 times higher than for BEVs, reflecting regional heterogeneity. At the city level, EV energy demand was clustered in major economic cities having strong policy support and industrial infrastructure.[26] Cities in the Yangtze River Delta region, such as Shanghai and Hangzhou, dominated EV demand. Shenzhen and Guangzhou combine strong production bases with large consumer markets, while key cities in central and western China, such as Chengdu and Chongqing, have experienced rapid expansion.

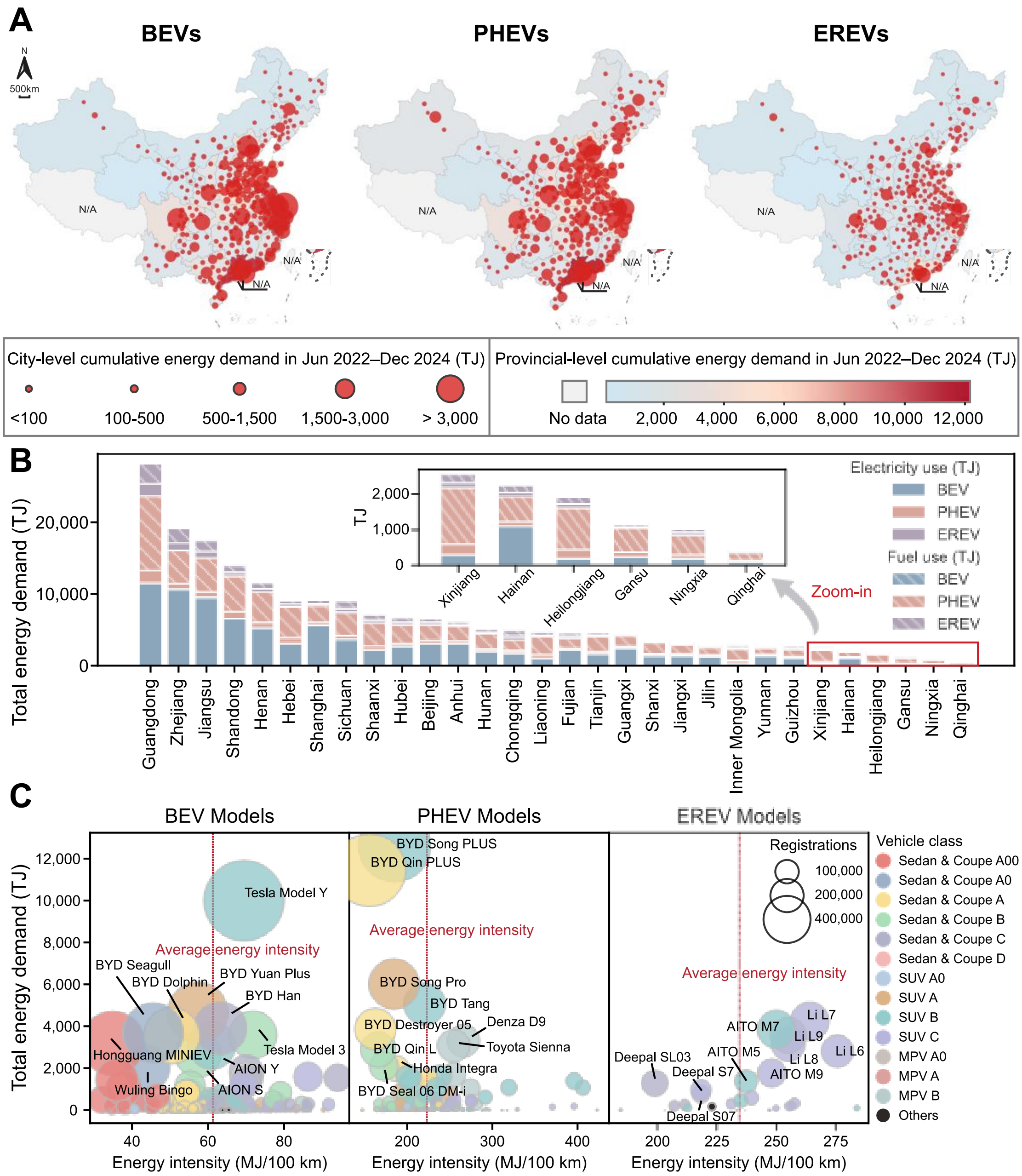

**Figure 4.** City-level cumulative energy demand of BEVs, PHEVs, and EREVs, June 2022–December 2024. (A) Geographical distribution of city-level energy demand by powertrain; (B) provincial-level energy demand sourced from electricity and gasoline among EV powertrains; and (C) energy demand of EV models by powertrain, intensity, and number of registrations.

Figure 4B illustrates provincial EV energy demand from 2022 to 2024 by source, underscoring the limited decarbonization progress made by China's EV fleet. Despite rapid electrification and strong EV market expansion, overall energy use remained ongoing reliance on gasoline for hybridized powertrains. In Guangdong, the largest EV market in China, fuel consumption (13,114 TJ) was nearly comparable to electricity demand (15,075

TJ) from mid-2022 to the end of 2024. Regional disparities further reinforced this trend, in which northeastern provinces such as Heilongjiang, as well as lagging EV regions such as Xinjiang showed a particularly high dependence on gasoline.

As illustrated in Figure 4C, EV energy demand from 2022 to 2024 was focused on a small number of best-selling models. For BEVs, Tesla Model Y and BYD Seagull accounted for the majority of consumption. PHEV demand was dominated by BYD Song PLUS, while EREV demand was driven primarily by large SUVs from Li Auto and AITO. These results highlight the strong influence of leading automakers in shaping real-world energy outcomes. Detailed policy implications are provided in Supplemental Note S2.

## DISCUSSION

In this section we discuss the robustness of our framework for real-world assessment of EV emissions. We modify the framework to account for electricity used for charging, then project the emissions outlook for China's EVs to 2035 under three scenarios.

### Robustness of the framework

The RF-based energy intensity estimation models achieved high predictive accuracy and strong generalizability (see Table S1), as confirmed by a grid search using 5-fold cross-validation hyperparameter tuning, as well as interpretability analysis of the selected key features based on the SHapley Additive exPlanations approach (SHAP)[27] (see Figure S7). For BEVs, eight key parameters, most prominently gross vehicle weight and battery capacity, emerged as dominant features, whereas PHEVs and EREVs exhibited more complex dependencies involving more than 25 technical features. We further evaluated sensitivity to potential sampling bias by conducting repeated reduced-sample training at 50%, 70%, and 90% of the available training data. Model performance remained stable across sampling ratios, with mean variations in held-out test $R^2$ of 0.03 for BEVs, 0.08 for PHEVs, and 0.09 for EREVs. These results suggest that the estimates are not unduly sensitive to moderate reductions in training data coverage and are therefore reasonably robust to potential sampling bias (see Figure S8).

To account for utilization heterogeneity, we applied a utilization correction and conducted a province-specific sensitivity analysis (see Figure S9), which reduced the national EV energy demand estimate by 10.2%. Historical EV emissions remained robust after a cumulative uncertainty analysis incorporating registration preprocessing, RF model prediction error, uncertainty in the share of charging electricity for hybrid EVs, and electricity emission factor uncertainty (see Figure S10).

Our real-world energy intensity estimates closely align with IEA benchmarks[c]: the fleet-average BEV intensity differed by only 1.68 MJ/100 km from the benchmark for China. After incorporating real-world electricity emission factors, fuel efficiency, and the observed shares of electricity use for charging, the estimated carbon intensities averaged 92.3 $gCO_2$/km per vehicle for BEVs—17.5 $gCO_2$/km above the IEA benchmark[d]—while PHEVs and EREVs reached 194.8 and 217.2 $gCO_2$/km per vehicle, exceeding the IEA estimates by 58.6 and 81 $gCO_2$/km, respectively. As an external plausibility check, we further compared our city-level electricity demand estimates with an independent public charging-order dataset for three major cities,[28] which shows consistent cross-city ranking and close agreement in aggregate values (see Figure S11).

---

[c] IEA (2024). Global EV Outlook 2024. IEA, Paris. https://www.iea.org/reports/global-ev-outlook-2024.

**Modified framework considering charging electricity**

Our modified accounting framework for PHEVs and EREVs calculated emissions from electricity and gasoline separately following China's GB/T 19753-2021 standard. For each model, we systematically combined real-world electricity and gasoline intensities with the observed shares of electricity use and fuel consumption to estimate annual vehicle kilometers traveled under each energy source. This approach captures how drivers actually utilize electric and combustion modes, rather than relying on simplified assumptions such as a fixed UF (see Figure S12 and Supplemental Note S3).

**Emissions outlook under three scenarios**

Using historical emissions data for China's EV fleet from 2016 to 2024, we projected operational $CO_2$ emissions to 2035 under three scenarios: the Roadmap 2.0 benchmark, the business-as-usual (BAU) scenario, and the trend-and-policy scenario (see Figure 5). As shown in Figure 5A, by 2024 total EV emissions have reached 14.2 megatonnes of $CO_2$ ($MtCO_2$), already exceeding the 2035 target of Roadmap 2.0 by approximately 4 $MtCO_2$. PHEVs and EREVs accounted for much of the growth, as their YoY emissions rose by 84.7% in 2023 and 91.7% in 2024 (see Figure 5C), roughly four times faster than BEV emissions (see Figure 5B).

Monte Carlo projections diverged markedly across scenarios. As shown in Figure 5C, the Roadmap 2.0 scenario projects PHEV and EREV emissions of only 0–0.78 $MtCO_2$ in

[d] IEA (2025). Global EV Data Explorer. IEA, Paris. https://www.iea.org/data-and-statistics/data-tools/global-ev-data-explorer.

2035, a pathway that appears increasingly inconsistent with observed market dynamics.[29] Under the BAU scenario, total EV emissions would continue to rise without peaking before 2035, reaching 17.0–26.1 $MtCO_2$. Within this total, BEVs would contribute 9.4–17.6 $MtCO_2$, while PHEVs and EREVs would contribute 7.5–8.5 $MtCO_2$. By contrast, under the trend-and-policy scenario, total EV emissions would peak around 2030 at 21.1–30.9 $MtCO_2$. Emissions from PHEVs and EREVs would also peak around 2030, at 10.1–14.0 $MtCO_2$, before declining thereafter.

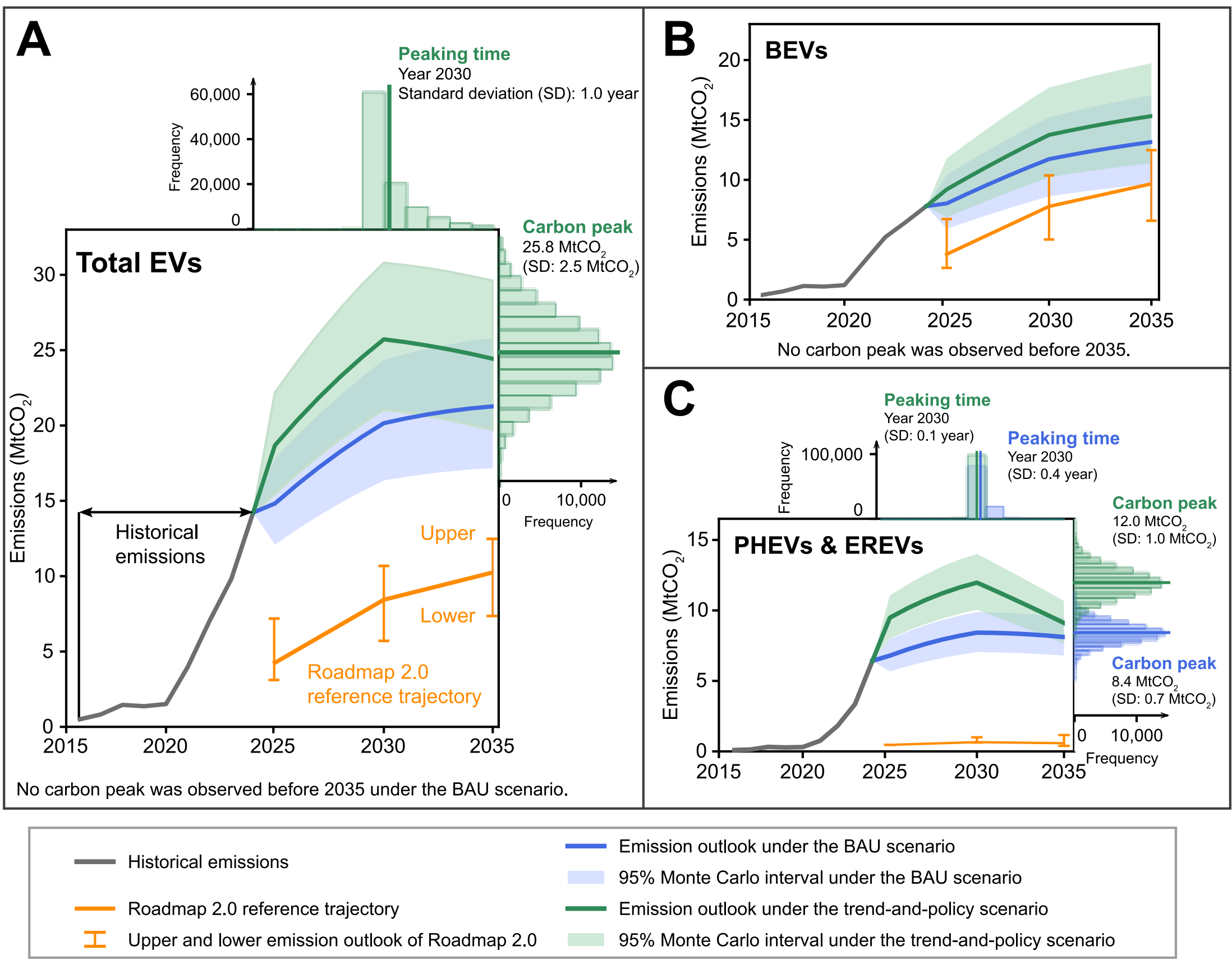

**Figure 5.** Monte Carlo projections of operational $CO_2$ emissions from China's EV fleet to 2035 under the Roadmap 2.0 benchmark, the BAU scenario, and the trend-and-policy scenario: (A) total EVs; (B) BEVs; and (C) PHEVs and EREVs.

Sensitivity analyses further quantify how uncertainty in key assumptions affects both the magnitude and timing of the carbon peak (see Figure S13). Under the trend-and-policy scenario, variation in the total EV carbon peak would be dominated by the electricity emission factor (59.0%), followed by BEV sales (17.9%), electricity intensity (11.5%), and PHEV/EREV sales (10.7%). In contrast, variation in peaking time would be driven primarily by PHEV/EREV sales (43.5%), BEV sales (35.0%), and the electricity emission factor (12.8%). For PHEVs and EREVs specifically, sales uncertainty would remain the dominant driver of the carbon peak across scenarios, while peaking time would depend mainly on sales, UF, and the electricity emission factor.

In this study, we developed a city-level, bottom-up framework to assess the real-world energy demand and WTW emissions of China's EV fleet, integrating powertrain heterogeneity with spatially explicit carbon intensities. Several limitations remain, however. First, the current analysis was limited to WTW energy demand and $CO_2$ emissions and did not include full life-cycle factors such as vehicle manufacturing, recycling, or city-level life cycle assessments. Future work should broaden the analytical boundary to incorporate vehicle stock turnover. Second, our analysis aggregated annual vehicle kilometers traveled at the regional level, which may overlook temporal dynamics in charging behavior, such as seasonal variation, policy shifts, and infrastructure expansion. Future work should incorporate model-specific and city-level user data, for example through vehicle trajectory records, to capture these dynamics more accurately and strengthen the policy relevance of EV emissions assessments.

## CONCLUSIONS

Using more than 20 million registrations covering 586 EV models across 295 cities in the 2020s, we constructed the first high-resolution, real-world database of China's EV market and projected transition pathways to 2035. Our key findings and associated implications[e] are summarized below.

- **A high-accuracy data-driven regression model bridged the gap between test-cycle and real-world data, affording the first comprehensive database of real-world energy intensity for more than 580 EV models.** Real-world intensities substantially exceeded test-cycle values, by 137.0% for EREVs, 97.7% for PHEVs, and 28.9% for BEVs. Among powertrains, BEVs exhibited the lowest real-world intensity, outperforming PHEVs and EREVs, and greatly exceeding the efficiency of ICEVs. High model accuracy provided strong generalizability to unseen testing data.

[e] Considering space limitations, the detailed policy implications are provided in Supplemental Note S2.

The results show that the unified regression framework provides a robust foundation for completing real-world datasets.

- **City-level emission assessments of more than 20 million EV registrations revealed the uneven advancement of China's EV transition, characterized by pronounced spatial and technological heterogeneity and shaped by EV powertrain composition, grid carbon intensity, and charging behaviors.** From 2022–2024, the average BEV carbon intensity was 88.5 g$CO_2$/km, significantly lower than PHEVs (207.6 g$CO_2$/km) and EREVs (226.0 g$CO_2$/km). Gasoline still accounted for 44% of total EV energy use nationwide, underscoring the urgency of grid cleaning, regional policy alignment, and stronger automaker commitments to reduce fossil fuel dependence.
- **EV-related operational emissions reached 14.2 Mt $CO_2$ in 2024 amid rapid market expansion. Under a trend-and-policy pathway, emissions would peak around 2030 and then decline, driven by accelerated shifts from PHEVs and EREVs to more efficient BEVs, supported by technological breakthroughs and strengthened policy alignment.** Notably, China's EV fleet emissions already exceeded the Roadmap 2.0 benchmark for 2035 by approximately 4 Mt $CO_2$, largely driven by PHEV and EREV emissions, which increased by 84.7% in 2023 and 91.7% in 2024, about four times faster than BEV emissions. Under the BAU scenario, EV emissions would continue rising, reaching 17.0–26.1 Mt $CO_2$ by 2035. In contrast, the trend-and-policy scenario would produce a transitional peak around 2030 (21.1–30.9 Mt $CO_2$), followed by a decline to 19.8–29.6 Mt $CO_2$ by 2035, underscoring that coordinated policy alignment and technological innovation are critical for achieving long-term decarbonization.

# EXPERIMENTAL PROCEDURES

## Resources availability

### *Lead contact*

Requests for further information and resources should be directed to and will be fulfilled by the lead contact, Dr. Minda Ma (minda.ma@cqu.edu.cn).

### *Data and code availability*

Any additional information required to reanalyze the data reported in this paper is available from the lead contact upon request.

### *Materials availability*

This study did not generate new unique materials.

## Estimating real-world energy intensities

We developed a unified RF-based regression framework to estimate the real-world energy intensities of BEV, PHEV, and EREV models. Using crowd-sourced empirical samples collected under real-world driving conditions (see Supplemental Method S1) and technical vehicle feature data (see Table S3), we conducted data preprocessing (see Supplemental Method S2) and classified each powertrain type into vehicle class segments according to Table S4. Then, we constructed separate RF regressors $\mathcal{F}^{(k)}$ for each powertrain type $k$. For each $k$, the dataset was randomly divided into training (80%) and test (20%) subsets. Each model was trained on the training set with embedded feature selection and five-fold cross-validated hyperparameter optimization. Out-of-sample predictive accuracy and model interpretability were evaluated on the held-out test set using four metrics (see Table S5) and SHAP analysis, respectively. For vehicle models lacking empirical samples, real-world energy intensities were predicted using the optimized RF model, $\hat{\mathcal{F}}_{RF}^{(k)}$. Detailed model specifications and validation procedures are provided in Supplemental Method S3.

### Accounting framework

Using the estimated real-world energy intensities, we developed a city-level, bottom-up framework to quantify WTW energy demand and $CO_2$ emissions from EV powertrains. As defined in Supplemental Notes S4–S5, BEV emissions arise entirely from electricity generation, whereas PHEVs and EREVs involve emissions from both electricity generation and gasoline combustion. Accordingly, we developed separate accounting models for BEVs and for PHEVs/EREVs, as detailed in Supplemental Methods S4–S6.

### Scenario design

Building on historical emissions data from 2016 to 2024, we projected China's EV transition pathways to 2035 under three market penetration scenarios: the Roadmap 2.0 scenario[f], the BAU scenario, and the trend-and-policy scenario[g].[30,31] These scenarios were defined by overall EV penetration levels and the relative shares of BEVs, PHEVs, and EREVs in fleet evolution (see Supplemental Method S7 and Table S6). To characterize uncertainty in key assumptions (Table S7)—including BEV, PHEV, and EREV sales, electricity emission factors, electricity intensity, gasoline intensity, and UF—we conducted 100,000 Monte Carlo simulations for the BAU and trend-and-policy scenarios. These simulations were used to estimate the 95% uncertainty intervals of annual emissions, peak timing, and peak emissions. In addition, Spearman rank correlation analysis was applied to quantify parameter sensitivities for carbon peak and peaking time. Detailed methods are provided in Supplemental Method S8.

## SUPPLEMENTAL INFORMATION

The supplemental materials, which comprise text, figures, tables, and references, are provided at the end of this submission file.

[f] China Society of Automotive Engineers. (2020). Energy-saving and New Energy Vehicle Technology Roadmap, 2.0 Edition (China Machine Press). https://en.sae-china.org/a3967.html.

[g] IEA (2024). Global Energy and Climate Model. IEA, Paris. https://www.iea.org/reports/global-energy-and-climate-model.

## ACKNOWLEDGMENTS

The first author appreciates the Chongqing Association for Science and Technology Think Tank Research Project (2025KXKT40). Contributions from Dr. Nan Zhou was supported by the Energy Demand Changes Induced by Technological 30 and the Social Innovations network, an initiative coordinated by the Research Institute of Innovative Technology for the Earth and the International Institute for Applied Systems Analysis and funded by the Ministry of Economy, Trade, and Industry, Japan.

## AUTHOR CONTRIBUTIONS

Conceptualization, Y.D., M.M., and N.Z.; methodology, Y.D., H.Y., and X.M.; investigation, Y.D., M.M., H.Y., and Z.M.; writing—original draft, Y.D. and M.M.; writing—review & editing, Y.D., M.M., and N.Z.; funding acquisition, M.M., and N.Z.; supervision, M.M. and N.Z.

## DECLARRATION OF INTERESTS

None.

## FIGURE TITLES AND LEGENDS

**Figure 1.** Distribution of EV registrations in China, Jun 2022–Dec 2024. (A) Provincial EV registrations and proportions of BEVs, PHEVs, and EREVs; (B) monthly registrations and YoY growth trends; and (C) registrations by vehicle class segments for various periods.

**Figure 2.** Real-world energy intensity (EI) distributions and estimates based on RF regression. (A) Empirically derived energy intensity distributions for samples of BEV, PHEV, and EREV models; (B) model fitting results via RF regression; and (C) overall energy intensity distributions of all BEV, PHEV, and EREV class segments after combining empirical samples and RF model predictions.

**Figure 3.** Real-world average carbon intensities of EVs in China (2022–2024). (A) Provincial heterogeneity of carbon intensities for BEVs, PHEVs, and EREVs; and (B) carbon intensities

generated by electricity and gasoline among vehicle class segments. Note: “Others” shown in Panel B include light vans, microvans, and pickup trucks.

**Figure 4.** City-level cumulative energy demand of BEVs, PHEVs, and EREVs, June 2022–December 2024. (A) Geographical distribution of city-level energy demand by powertrain; (B) provincial-level energy demand sourced from electricity and gasoline among EV powertrains; and (C) energy demand of EV models by powertrain, intensity, and number of registrations.

**Figure 5.** Monte Carlo projections of operational $CO_2$ emissions from China’s EV fleet to 2035 under the Roadmap 2.0 benchmark, the BAU scenario, and the trend-and-policy scenario: (A) total EVs; (B) BEVs; and (C) PHEVs and EREVs.

## FORMULAE AND EQUATIONS

None.

## TABLE TITLES AND LEGENDS

None.

## STANDARDIZED DATA REPORTING

None.

## SUPPLEMENTAL INFORMATION DESCRIPTION

The supplemental information includes one PDF file.